\documentstyle[psfig,frontieres,art10]{article}
\begin{document}

\heading{%
%Begin Heading 
% 
Gravitational event search with five resonant antennas % 
%End Heading 
}

\author{Sergio Frasca$^{1,2}$}

\address{%First address 
Diparimento di Fisica Universit\'a "La Sapienza", I-00185 Rome  
}
\address{%Second Address
INFN Roma 1
}

\begin{abstract}
Five cryogenic resonant gravitational antennas are now in operation.

This is the first time that such a large number of high sensitive antennas
are taking data and an agreement on data exchange has been signed by the
responsible groups.

The data exchanged will consist essentially in lists of ''candidate events''.

In this paper the procedure used by the Rome group in order to obtain
''candidate events'' is presented.

Some methods of analyzing the data of the "network" of the five antennas are
shown.
\end{abstract}

\section{The International Gravitational Event Collaboration}

In table I there is a list of the five cryogenic resonant antennas now in
operation. These detectors are aluminum (niobium in the case of the
Australian one) cylindrical bars, all equipped with resonant transducers,
that give them two narrow bands of detection of a few hertz around the two
near coupled modes of the bar-transducer system, at frequencies of about 900
Hz (about 700 Hz for Niobe). These are called the detection bands of the
antenna; in some detectors they can constitute a single wider band.

The orientation of the bars was chosen in order to achieve the maximum
parallelism of the bars.

In order to coordinate and correlate the data produced by these detectors,
in July 1997, all the groups that operate them have undersigned a data
exchange protocol named ''International Gravitational Event Center''
(''IGEC''). It is based on some technical and policy rules. The main points
are:

- The data consist of ''candidate events'' and service information.

- The minimum information about each event consist of the time of the event
maximum, the amplitude in units of standard burst strain, the duration and
the mean detector noise at the time of the event.

- Additional fields, giving further information on the candidate events
(e.g. the shape and/or parameters relating the shape) may be provided.

- Time accuracy will be at least 0.1 s.

- Each group will set up a site (ftp, www,...) on which the data will be
continuously available in agreed formats, with an updating rate not
exceeding one day.

The full text of the agreement, together with other data exchange documents,
is posted at {\it http://grwav1.roma1.infn.it/gwda/de/de.htm} .

In this paper I will show how the candidate events are produced by the Rome
group (that is operating the two cryogenic resonant antennas Explorer and
Nautilus) and how the IGEC data can be analyzed.

\[
%%%%pia\stackrel{\text{{\large Table 1}}}
{
\begin{tabular}{|l|l|l|}
\hline
{\bf Antenna} & {\bf Institution} & {\bf City} \\ \hline
Allegro & LSU & Baton Rouge \\ \hline
Auriga & LNL (INFN-AURIGA) & Legnaro \\ \hline
Explorer & CERN (INFN-ROG) & Geneva \\ \hline
Nautilus & LNF (INFN-ROG) & Frascati \\ \hline
N\"{i}obe & UWA & Perth \\ \hline
\end{tabular}
} 
\]

\section{How the events are obtained}

The simplified modelisation of the detector as a linear system is normally
very good and the gravitational signal is simply added to the noise;
nevertheless the detection of short gravitational pulses (with unknown
shape) in the data of gravitational wave antennas is not an easy task. The
main reasons are:

- the low signal-to-noise ratio (SNR) and the rarity of the expected pulses

- the ignorance of their shape

- the non-stationarity of the noise of the detectors and the presence of
many spurious events.

As regards the shape of the pulses, resonant antenna people normally
consider as a standard pulse a delta function (or the part of that signal in
the detector band) that, in the data at the output of the antenna, becomes a
known waveform. The noise is considered gaussian and stationary (also if
some consider it slowly non-stationary), completely described by the noise
power spectrum.

With these assumptions, the problem becomes the classical problem of
detecting a known waveform in gaussian noise, that is optimally solved by
the ''matched filter''. This filter is a non-causal linear system that has
the property that at its output the signal-to-noise ratio (i.e. the ratio
between the square of the maximum of the signal and the variance of the
noise) is maximized.

This means that, if we normalize the filter in order to have the same
amplitude of the maximum of the input and output signal, the output noise
(that is also gaussian because of the linearity of the filter), has the
minimum variance, so the probability that the noise samples exceed the
threshold is minimized.

There are many ways to express the equation of the matched filter. In the
frequency domain it is

\begin{equation}
M(j\omega )=\frac{F^{*}(j\omega )}{S(\omega )}  \label{mf}
\end{equation}
where $F^{*}(j\omega )$ is the complex conjugate of the Fourier transform of
the response of the antenna to the standard pulse and $S(\omega )$ is the
noise power spectrum.

As a matter of fact, each group implement practically the filter in one or
more different ways. This because:

- there are different ways of taking data (high frequency sampling, aliased
sampling, two lock-ins acquisition, single central lock-in,...)

- the use of frequency domain procedures or time domain procedures

- the use of adaptive or non-adaptive procedures

- the threshold mechanism

- particular procedures of event features estimation and/or spurious event
detection.

In our case (Ref \cite{fmp}), the most advanced procedure that we use is
based on a high frequency sampling (5000 Hz), from which we extract a band
of 40 Hz centered at 816 Hz, containing the detection bands.

Other bands are also analyzed in order to monitor the behavior of the
antenna and the presence of local disturbances, but this will not be
discussed here.

On these data we implement three adaptive matched filters (and also filters
of other types), followed by a two state adaptive threshold mechanism.

\section{The adaptive matched filters}

The noise power spectrum of our antennas is not stationary: both the
electric noise, that normally gives a wide-band and a narrow band
contributions, and the seismic noise, that gives normally a narrow band
contribution, change with time. Moreover often extra noise peaks appear in
the spectrum, due to local disturbances. These bands change slowly in
amplitude and in frequency. Almost all these non-stationarities have time
constants of at least a few hours. The presence of the non-stationarities
can be seen as the fact that the detector has a time-varying sensitivity.

Also the resonance frequencies of the antennas can slowly change, mainly
because of the change in the transducer polarization voltage. In order to
have a good filter in presence of these varying features of the detectors,
we use an adaptive approach: we apply the filter in the frequency domain,
using the expression of eq.\ref{mf} and use, for the power spectrum, a
first-order auto-regressive sum of the periodograms, with a time constant of
one hour and with an updating rate of about one minute. The filter is so
recomputed about every minute; this is our ''basic'' adaptive filter.

In presence of a big disturbance, the basic filter estimated spectrum is
''dirtied'' and can remain dirty for a long time (also many time constants)
after the end of the disturbance, producing filters that are not optimal for
the ''clean'' data. To solve this problem, we have implemented a matched
filter that uses only ''clean'' periodograms for the estimation of the power
spectrum.

We studied also another policy that gets good results in the case that the
disturbed period can be seen as a low sensitivity period. It is based on the
reduction of the auto-regressive time constant in the case of highly
disturbed spectra.

In order to evaluate the performances of the different filters, we add to
the acquired data fictitious ''theoretical'' pulses and analyze the
resulting SNR; this is done every half an hour, disturbing the real data for
less than the $0.1\%$ of the time. We chose as ''official filter'' the
filter that gives the better results at that time. This procedure provides
also a sort of Monte Carlo check procedure for the whole system and can be
used also to analyze the behavior in case of non-delta-pulse events.

\section{The threshold mechanism}

\subsection{The adaptive threshold}

Because of the non-stationarity of the antenna noise, also at the output of
one matched filter (or of any linear filter) the noise is not stationary,
i.e. the variance of the noise changes (slowly) with the time; this means
that the sensitivity of the detector changes with the time. If we put a
fixed threshold on these data, we have more (spurious) candidate events when
higher is the noise (and lower is the sensitivities) and this can highly
worsen the statistics. So we must change the threshold with the time.

This can be done in various ways, for example choosing a fixed number N and
taking the N highest events of each hour.

We use a different procedure, based on an ''adaptive threshold'' defined in
the following way.

Let $x_{i}$ be the filtered data samples. We estimate the background
statistics by computing the auto-regressive mean of the absolute value and
of the square of $x_{i}$

\begin{equation}
m_{i}=(1-w)\cdot |x_{i}|+w\cdot m_{i-1}  \label{mi}
\end{equation}

\begin{equation}
q_{i}=(1-w)\cdot x_{i}^{2}+w\cdot q_{i-1}  \label{qi}
\end{equation}
with 
\begin{equation}
w=e^{-\frac{\Delta t}{\tau }}  \label{w}
\end{equation}
where $\Delta t$ is the sampling time and $\tau $ is the ''memory'' of the
auto-regressive mean (we normally chose $\tau =600\ s$).

Then we define the standard deviation 
\begin{equation}
\sigma _{i}=\sqrt{q_{i}-m_{i}^{2}}  \label{sigi}
\end{equation}
and the threshold $\theta $ is not set on $|x_{i}|$, but on the critical
ratio of $|x_{i}|$ given by 
\begin{equation}
z_{i}=\frac{|x_{i}|-m_{i}}{\sigma _{i}}  \label{thr}
\end{equation}

This procedure was developed for the general case of $x_{i}$; if $x_{i}$ is
simply zero mean, as is in the case of the matched filter, the algorithm for
estimating the adaptive threshold can be simplified, using just the
estimation $q_{i}$ and computing the critical ratio as $z_{i}=|x_{i}|/\sqrt{%
q_{i}}$.

In case of very large value of $\sigma _{i}$, we can reduce the memory $\tau 
$ in order to reduce the ''blinding'' effect at the end of large
disturbances.

\subsection{The event definition: the two state event machine}

In order to define candidate events, let us suppose that we have chosen an
adaptive threshold $\theta$ and a dead time (that is the minimum time
between two different events; it depends on the apparatus, the noise and the
expected signal: we normally use 3 s).

Then we use an easy two-state (0 and 1) mechanism that we call the ''event
machine''. The algorithm performed is the following.

- The machine is normally set in the state $0$.

- When the signal goes over the threshold, it changes to state $1$ and an
event begins.

- The state changes to $0$ after the signal has remained below the threshold
for a time longer than the dead time.

- The ''duration'' of the event is the duration of the state $1$, subtracted
the dead time.

A simplified model of this algorithm (Ref \cite{fp}) is a two-state Markov
chain (in the discrete time). Its transition matrix is 
\begin{equation}
\left( 
\begin{array}{cc}
1-p_{01} & p_{01} \\ 
p_{10} & 1-p_{10}
\end{array}
\right)   \label{mc}
\end{equation}
where $p_{01}$ and $p_{10}$ are the transition probability for the two
states. We can easily compute the probabilities of the two states as 
\begin{equation}
p_{0}=\frac{p_{10}}{p_{01}+p_{10}}  \label{p0}
\end{equation}
\begin{equation}
p_{1}=1-p_{0}  \label{p1}
\end{equation}
and the mean length (in unit of sampling time) 
\begin{equation}
\overline{L}=\frac{1}{p_{10}}  \label{l}
\end{equation}

\subsection{The event density}

We can define the event density $\lambda $ as the number of the events per
unit time and then the candidate events production can be modeled by a
Poisson process with parameter $\lambda $.

Obviously the value of $\lambda $ depends on the value of the adaptive
threshold $\theta $. The functional dependence of $\lambda $ on $\theta $
depends on the distribution of the filtered data (or, more precisely, on its
tail), that only theoretically is gaussian; in practice it depends strongly
on the local disturbances.

This is what finally limits the sensitivity of a single antenna. However,
because the disturbances give heavy tails in the distributions, reducing the
threshold (and so enhancing the sensitivity) in this region gives a not big
increment on $\lambda $. It is not so for the gaussian distribution (that
has light tails).

\section{The coincidences}

If we consider the case of $N$ antennas, each defined with a $\lambda _{i}$,
and if we choose a coincidence window of duration $t_{w}$, in the hypothesis
of uncorrelated data, that is that of the background noise event, we have an
expected density of ''casual'' coincidences given by 
\begin{equation}
\lambda ^{(N)}=t_{w}^{N-1}\cdot \prod_{i=1}^{N}\lambda _{i}  \label{lam}
\end{equation}
$t_{w}$ is chosen depending on the apparatuses, the time precision and the
light time delay between the antennas; good values ranges between $0.1$ and $%
1\;s$. In table 2 there are some values for the coincidence densities
(expressed in number of coincidences per day) for the case that all $\lambda
_{i}$'s are equal to the value $\lambda $ and $t_{w}=0.1\;s$.

\[
%%%%%%%%pia\stackrel{\text{{\large Table 2}}}
{
\begin{tabular}{|c|c|c|}
\hline
& ${\bf \lambda =100\;events/day}$ & ${\bf \lambda =1000\;events/day}$ \\ 
\hline
${\bf N=2}$ & $1.15\cdot 10^{-2}$ & $1.15$ \\ \hline
${\bf N=3}$ & $1.34\cdot 10^{-6}$ & $1.34\cdot 10^{-3}$ \\ \hline
${\bf N=4}$ & $1.55\cdot 10^{-10}$ & $1.55\cdot 10^{-6}$ \\ \hline
${\bf N=5}$ & $1.79\cdot 10^{-14}$ & $1.79\cdot 10^{-9}$ \\ \hline
\end{tabular}
} 
\]

In order to evaluate the statistical significance of the found coincidences,
one should consider the appropriate Poisson statistics.

Anyway, the Poisson model is normally not a good approximation, because of
all the non-stationarities in the detection process. In practice the $%
\lambda $'s are functions of the time. In particular, there are

- periodicities (e.g. the solar day)

- aggregation (the disturbances often are clustered in time)

- holes in the data.

Also the signals (the ''true'' events) are not expected to be uniformly
distributed in time, because of the radiation pattern of the antennas and
the non-uniformity of the space distribution of the biggest sources, that
causes a sidereal day modulation of the detection probability.

To have an efficient evaluation method in presence of this problems, in the
case of two antennas, Weber introduced a non-parametric procedure, that we
call "pulse correlation".

\subsection{The pulse correlation}

The pulse correlation method is based on the evaluation of the casual
(''background'') coincidence rate for two antennas obtained by adding a bias
time $\tau $ to the eventts of one antenna. $\tau $ is normally set equal to 
$k\cdot t_{w}$, with $k$ integer and $-n\leq k\leq n$ (e.g., $n$ can be set
equal to 1000).

Taken a period of time $T\,$ (e.g. one day, one month,...) the number of
coincidences $C(\tau )$ are computed. If we take the mean value $\mu _{C}$
of $C(\tau )$ excluding $C(0)$, the estimated number of ''true''
coincidences is given by 
\begin{equation}
\widetilde{C}=C(0)-\mu _{C}  \label{c0}
\end{equation}
if $\widetilde{C}\geq 0$. We can define $\lambda _{C}=\mu _{C}/T\,$ as the
estimated chance coincident event density (analogous to $\lambda ^{(2)}$ for
the equation \ref{lam}).

Because of the non-stationarities, the expected shape of $C(\tau )$ for $%
\tau \neq 0$ is not uniform and the evaluation of $\mu _{C}$ deserves some
cares.

I would note the analogy between the pulse correlation function and the
classical cross-correlation function.

To evaluate the chance probability of having $C(0)$, one can use one of the
two following methods:

a) use the Poisson statistics with parameter $\mu _{C}$

b) compute the histogram of $C(\tau )$ and evaluate the ''frequentistic''
probability of having $C(0)$.

Both methods must be used with care, the first because of the uncertainty in
the evaluation of $\mu _{C}$ and the second because it assumes that the
event production is stationary on times of the order of $n\cdot t_{w}$.

A particular use of the pulse correlation is in correlating the events of an
antenna with themselves (''pulse auto-correlation''). This method of
analysis can be used to identify the non-stationarities of the events: in
particular aggregation and periodicities.

The pulse correlation can be applied on data selected by some ''single
event'' rules (e.g. the amplitude or the time of occurrence, that can be a
particular solar or sidereal hour) or ''coincidence'' rules (e.g. consider a
coincidence only if the amplitude and/or the duration of the two events is
about the same).

In the evaluation of the probabilities, particular care must be taken on any
choice made ''a posteriori''.

\subsection{The pulse correlation in the case of more than two antennas}

How can we generalize the pulse correlation in the case of $N>2$ antennas ?
Consider the following two methods:

a) the {\bf couple pulse correlation, }obtained summing the pulse
correlation of all the $N\cdot (N-1)/2$ couples of antennas. We obtain a
function $C_{N}(\tau )$. With these method there is a ''natural'' weighting
of multiple coincidences. In fact a coincidence between only two antennas is
considered just once, a coincidence between 3 antennas is considered 3 times
, a 4 antennas coincidence it is considered 6 times and a 5 antennas
coincidence is considered 10 times. The background is the sum of all the
backgrounds.

b) the {\bf multiple pulse correlation }$C(\tau _{1},\tau _{2},...,\tau
_{N-1})${\bf , }obtained by adding to the events of each of the first $N-1$
antennas a different bias time. The delay variables $\tau _{i}$ can be
chosen equal to $k\cdot t_{w}$, with $-n\leq k\leq n$ (e.g., $n$ can be set
equal to 10; in this case, with five antennas, we have 194481 delays).

In searching for coincidences for more then two antennas, it must be taken
into account that, in case of not too much big gravitational events, the
probability that an event is overlooked by a detector is not negligible.
This because of the presence of the additive noise (see, for example, Ref 
\cite{app}), the difference in the frequencies, the not perfect parallelism
and the difference in sensitivities, that, also if are basically similar,
change in time depending on the local noises. So methods like the multiple
pulse correlation are good only in case of huge events and, for small
events, would produce a false dismissal probability near to 1.

More complex procedures were presented in Ref \cite{fp}. With these
procedures, with five detectors it is possible to estimate also the position
of the source in the sky and the polarization of the gravitational pulse,
together with a better estimate of its energy. Also the rejection of
spurious events is enhanced. Unfortunately these methods get poor results
with parallel antennas.

\section{Another perspective}

Because of the rarity of the gravitational events, it is important to not
overlook the presence of real events in the data. So any method used to
reduce the false detection probability, should not enhance too much the
false dismissal probability. This means that we should reject a candidate
event only after a careful analysis.

On the other hand, also very low probability coincidence events must be
carefully checked for consistency, e.g. to compare the amplitudes and the
lengths in the different antennas or to check in the other antennas, that
did not detect events at that time, if there is ''something'' (i.e. an event
under the threshold).

As it is shown in table 2, the multiple coincidence operation (2 or 3
antennas on 5) reduces strongly the number of candidate events; this can be
done automatically. For the surviving candidates we should analyze carefully
(i.e. not automatically) the outputs of all the antennas in operation, the
signal shapes and the auxiliary channels.

It is important to have a data-base of all the events, with some
characteristics (e.g. the "color", i.e. some spectral parameters, the
length, the shape,...).

A very important information to take into account are other impulsive
astrophysical events, e.g. the gamma ray bursts and the neutrinos bursts
detected in underground experiments. These must constitute a complementary
data base to be analyzed routinely together with the candidate events. Also
supernova surveys can be useful, in order to correctly interpret the other
detector results, also if they don't give precise events time.

After all, the ''true decision'' on the ''promotion'' of a candidate event
will be made not by an algorithm, but by a long discussion.

\vfill 

\end{document}